\documentclass[12pt,epsfig]{article}
\newfont{\Bbb}{msbm10 scaled 1200}     %instead of eusb10
\newcommand{\mathbb}[1]{\mbox{\Bbb #1}}

\def\tr{{\rm tr}}

\def\lbldef#1#2{\expandafter\gdef\csname #1\endcsname {#2}}
\def\eqn#1#2{\lbldef{#1}{(\ref{#1})}%
\begin{equation} #2 \label{#1} \end{equation}}

\def\href#1#2{#2}

\newcommand{\beq}{\begin{equation}}
\newcommand{\eeq}{\end{equation}}
\newcommand{\ber}{\begin{eqnarray}}
\newcommand{\eer}{\end{eqnarray}}

\newcommand{\beqar}{\begin{eqnarray}}

\newcommand{\eeqar}{\end{eqnarray}}
\begin{document}
\baselineskip=15.5pt
\renewcommand{\theequation}{\arabic{section}.\arabic{equation}}
\pagestyle{plain}
\setcounter{page}{1}
%\renewcommand{\thefootnote}{\fnsymbol{footnote}}
%--------+---------+---------+---------+---------+---------+---------+
%Title page
\begin{titlepage}

\leftline{\tt hep-th/0104036}

\vskip -.8cm

\rightline{\small{\tt CALT-68-2325}}
\rightline{\small{\tt CITUSC/01-010}}  
\rightline{\small{\tt NSF-ITP-01-25}}

\begin{center}

\vskip 2 cm

{\Large {An Exact Solution to Seiberg-Witten Equation}}

\vskip .5cm

{\Large{of Noncommutative Gauge Theory}}  

\vskip 1.5cm
{\large Yuji Okawa$^1$ \ and \ Hirosi Ooguri$^{1,2}$}

\vskip 1.5cm

{$^1$ California Institute of Technology 452-48, Pasadena, CA 91125}

\smallskip 
{\tt okawa, ooguri@theory.caltech.edu}

\bigskip

{$^2$ Institute for Theoretical Physics, 
University of California, Santa Barbara, CA 93106}

\vskip 2cm

{\bf Abstract}
\end{center}

\noindent
We derive an exact expression for the Seiberg-Witten map
of noncommutative gauge theory. It is found by studying 
the coupling of the gauge field to 
the Ramond-Ramond potentials in string theory. Our result
also proves the earlier conjecture by Liu. 

\end{titlepage}

\newpage

%--------+---------+---------+---------+---------+---------+---------+
%Body

\section{Introduction}
\setcounter{equation}{0}

A noncommutative gauge theory can be realized by considering branes
in string theory with a constant NS-NS two-form field
\cite{Connes:1998cr}. 
In \cite{Seiberg}, it was shown that there are two equivalent 
descriptions of the theory, one in terms of ordinary 
gauge fields $A_i$ on a commutative space and another in terms 
of noncommutative gauge fields $\hat{A}_i$ on a noncommutative
space whose coordinates obey the commutation relation,
\eqn{thetaparameter}{ [ x^i, x^j ] = -i\theta^{ij}.}
The map between $A_i$ and $\hat{A}_i$, called the Seiberg-Witten
map, is characterized\footnote{As pointed out in \cite{Asakawa},
there is a possibility to modify the equation by
performing field redefinition and gauge transformation.} by
the differential equation with respect to $\theta$, 
\eqn{sw}
{ \delta \hat{A}_i(\theta) 
 = - {1 \over 4} \delta \theta^{jk} \left[
 \hat{A}_{j}* (\partial_{k} \hat{A}_i
 + \hat{F}_{ki}) +  (\partial_{k} \hat{A}_i
 + \hat{F}_{ki}) * \hat{A}_i \right],}
with the initial condition,
\eqn{initial}
{ \hat{A}_i(\theta=0) = A_i.}
Here $*$ is the standard star product,
\eqn{star}{ f(x) * g(x) = \lim_{y\rightarrow x}~
 \exp\left[ -i \theta^{ij} {\partial^2 \over
\partial x^i \partial y^j} \right] f(x) g(y) ,}
and the field strength $\hat{F}_{ij}$ is defined as
\eqn{fieldstrength}
 {\hat{F}_{ij} = \partial_i \hat{A}_j 
   - \partial_j \hat{A}_i + i \hat{A}_i * \hat{A}_j
-i \hat{A}_j * \hat{A}_i.}
The differential equation (\ref{sw}) is known as
the Seiberg-Witten equation.

There have been several attempts to solve the Seiberg-Witten
equation. In \cite{Okuyama}, it 
was pointed out that the map can be expressed 
in terms of a functional integral which quantizes the Poisson
structure $\tilde{\theta}^{ij}$ related to $\theta^{ij}$ by
\eqn{anothertheta}
{(\tilde{\theta}^{-1})_{ij} = (\theta^{-1})_{ij}
+ \partial_i A_j - \partial_j A_i.}
By perturbatively evaluating the functional integral, one can obtain
the Seiberg-Witten map order by order in a formal power series expansion
in $\theta$. In \cite{Jurco}, the Seiberg-Witten map
is expressed in terms of the Kontsevich 
map \cite{Kontsevich} which relates the star product associated with
$\theta^{ij}$ to the one associated with $\tilde{\theta}^{ij}$
given by (\ref{anothertheta}).\footnote{The method
developed in \cite{Jurco} is also applicable to
the case when $\theta^{ij}$ is not constant.}  
There is a procedure to compute the Kontsevich map
as a formal power series expansion. The two approaches  
are related to each other since the Kontsevich
map can be expressed in terms of a functional integral 
\cite{Felder} which is
similar to the one used in \cite{Okuyama}.

One can also try to solve (\ref{sw}) directly order by order
in a power series expansion.
The structure of the power series is
examined in \cite{Garousi}, \cite{Mehen}. It was shown that it
involves the so-called generalized star products, which
also appear in the expansion of the open Wilson line,
\eqn{open}
{  \int dx *\left[ e^{ikx} P\exp\left(i\int_0^1 
   \hat{A}_i(x+l\tau) l^i d\tau\right)\right],}
where 
\eqn{whatl}
{l^i = k_j \theta^{ji},}
and $*[\cdots]$ means that we take the standard star product
(\ref{star}) in the expansion of the expression in $[\cdots]$
in powers of $\hat{A}_i$. 
This suggests that the Seiberg-Witten map
can be expressed in terms of the open Wilson line. Based
on this observation and the earlier papers \cite{Okuyama},
\cite{Jurco} mentioned in the above paragraph, 
it was conjectured in \cite{Liu}
that the (inverse of) Seiberg-Witten map is given in the momentum space by
\ber \label{liu}
&& F_{ij}(k)  \equiv
 \int dx~ e^{ikx} \left(\partial_i A_j(x) - \partial_j A_i(x)
\right)\cr
&&= \int dx *\left[
e^{ikx} \sqrt{\det\left(1 - \hat{f}\theta\right)}
       \left({1 \over 1 - \hat{f}\theta} \hat{f}\right)_{ij}
P\exp\left(i\int_0^1 
   \hat{A}_i(x+l\tau) l^i d\tau\right)  \right],\cr
&&
\eer
where
\eqn{whatf}
{\hat{f}_{ij} = \int_0^1 \hat{F}_{ij}(x+l\tau) d\tau.}
Here we are using the same symbol $x$ to denote both the
commutative (in the first line) and the noncommutative 
coordinates (in the second line). 
The path-ordering with respect to $\tau$ is implicit
in this expression and throughout the rest of the paper. 
It is clear that (\ref{liu}) obeys the initial condition
(\ref{initial}). To the quadratic order in the power 
series expansion in $\theta$, it was also checked in \cite{Liu}
that (\ref{liu}) satisfies the Seiberg-Witten equation. 

\medskip

In this paper, we derive an exact expression for the Seiberg-Witten
map. We will discuss the case where the gauge group is $U(1)$. 
Solving the Seiberg-Witten equation is equivalent to finding a two-form
$F_{ij} = F_{ij}(\hat{A}_i;~ \theta)$ which
\medskip

\noindent
($a$) is gauge invariant,
\eqn{gaugeinv}{ F_{ij}(\hat{A}_i + \partial_i\hat{\lambda} + 
i \hat{A}_i * \hat{\lambda}
 -i \hat{\lambda} * \hat{A}_i;~\theta) = F_{ij}(\hat{A}_i;~\theta),}

\noindent
($b$) obeys the Bianchi identity for the ordinary gauge theory:
\eqn{bianchi}{ \partial_{i} F_{jk} + \partial_{j} F_{ki} 
+\partial_{k} F_{ij} = 0,}

\noindent
($c$) satisfies the initial condition,
\eqn{initialtwo}{ F_{ij}(\hat{A}_i;~\theta=0) = \partial_i \hat{A}_j 
- \partial_j \hat{A}_i.}

\noindent
Modulo freedom of field redefinition and gauge transformation, 
the conditions ($a$) and ($b$) are equivalent
to the Seiberg-Witten equation since
the Bianchi identity ($b$) means that $F_{ij}$ can be expressed
as $F_{ij} = \partial_{i} A_{j}- \partial_j A_i$ for some $A_i$ and  
the gauge invariance ($a$) guarantees that, under the noncommutative 
gauge transformation,
\eqn{ncymgauge}
{\hat{A}_i \rightarrow \hat{A}_i + \partial_i\hat{\lambda} + 
i \hat{A}_i * \hat{\lambda}
 -i \hat{\lambda} * \hat{A}_i,}
$A_i$ transforms as an ordinary gauge field,
\eqn{ordinarygauge}
{A_i \rightarrow A_i 
+ \partial_i \lambda,}
for some $\lambda$ which depends
on $\hat{\lambda}$ and $\hat{A}_i$. 
These are exactly the conditions from which the
Seiberg-Witten equation was derived 
\cite{Seiberg}. The importance
of the condition ($b$) in this context was 
stressed in \cite{Mehen}. 

If we realize the noncommutative gauge
theory on $p$-branes in string theory, the two-form 
$F_{ij}$ obeying the three conditions ($a$) -- ($c$) can be 
found by identifying the current coupled to the Ramond-Ramond 
potential $C^{(p-1)}$. The gauge invariance ($a$) is manifest if
we use the point-splitting regularization on the string
worldsheet, and the Bianchi identity ($b$) is the consequence
of the gauge invariance of the Ramond-Ramond potential,
\eqn{ramondgauge}{ C^{(p-1)} \rightarrow C^{(p-1)}
+ d\epsilon, }
where $\epsilon$ is an arbitrary $(p-2)$ form in the bulk. From
the resulting expression for $F_{ij}$, it is straightforward
to verify that the initial condition ($c$) is satisfied.
The fact that the initial condition is satisfied 
is presumably related to the topological nature of
the Ramond-Ramond coupling and the lack of $\alpha'$
corrections to it.\footnote{
This result is in contrast to the case of the energy-momentum
tensor studied in our earlier paper \cite{Okawa:2001sh}. 
There it was shown that the energy-momentum
tensor of the noncommutative theory derived from the coupling
to the bulk graviton does not reduce to the one in the ordinary
gauge theory in the limit $\theta \rightarrow 0$.}

When the noncommutative space is $2n$ dimensional, namely,
when the rank of $\theta$ is $2n$, the Seiberg-Witten 
map\footnote{
It is known that a solution to the Seiberg-Witten equation
is not unique. For example, there is the field redefinition
ambiguity we mentioned in the above. It would be interesting
to find out if this solution, which naturally comes from
the string theory computation, has a special status among all
possible solutions.}  
we find from the Ramond-Ramond current is 
\ber \label{oursolution}
&&F_{ij}(k)~~ + ~~\theta^{-1}_{ij}\delta(k)\cr
&=& \frac{1}{{\rm Pf}(\theta)}
\int dx * \left[ e^{ikx}
\left(\theta - \theta \hat{f} \theta\right)^{n-1}_{ij}
P\exp\left(i\int_0^1 
   \hat{A}_i(x+l\tau) l^i d\tau\right)  \right].
\eer
Here the integral $\int dx$
is over the space coordinates on the brane and is
normalized as
\eqn{integral}
{\int dx  = \int {dx^1 \cdots dx^{2n}\over (2\pi)^{2n}},} 
the two-form $(\theta - \theta \hat{f}\theta)^{n-1}_{ij}$ 
in the integrand is defined as
\ber \label{contraction}
&&
\left(\theta - \theta \hat{f} \theta\right)^{n-1}_{ij}\cr
&\equiv& -{1 \over 2^{n-1} (n-1)!}
\epsilon_{ij i_1 i_2 \cdots i_{2n-2}}
\cr
&&\times \int_0^1 d\tau_1
\left(\theta - \theta\hat{F}(x+l\tau_1)\theta\right)^{i_{1}i_{2}}
\cdots \int_0^1 d\tau_{n-1}
\left(\theta - \theta\hat{F}(x+l\tau_{n-1})
\theta\right)^{i_{2n-3}i_{2n-2}}, \cr
&&
\eer
and the pfaffian is normalized
as
\eqn{pfaffian}
{{\rm Pf}(\theta) = {1\over 2^n n!}
\epsilon_{i_1\cdots i_{2n}} \theta^{i_1i_2}
\cdots \theta^{i_{2n-1}i_{2n}}.}
Note that the right-hand side of
 (\ref{oursolution}) depends only on $\hat{A}_i(x)$,
$\theta^{ij}$ and $k_i$. In particular, the 
combination 
$(\theta-\theta \hat{f}\theta)^{n-1}_{ij}/{\rm Pf}(\theta)$
does not depend the normalization of the $\epsilon$-symbol.

In order to make the logical structure of this paper
transparent, we will first prove that (\ref{oursolution})
satisfies the three conditions ($a$) -- ($c$) independently
of the string theory origin of the formula. In particular,
the proof holds for any $n$ even though the string theory
computation only works for $n \leq 4$. After the
proof is completed, we will explain how the solution is
found from the string theory computation
of the Ramond-Ramond coupling.

It turns out that the map (\ref{oursolution})
can be re-expressed in the form (\ref{liu}). Thus
we have also proven the conjecture in \cite{Liu}. 
Since we now have the exact expression for the Seiberg-Witten
map, it may also be possible to find an expression 
for the Kontsevich map in the case of (\ref{anothertheta}).

This paper is organized as follows. 
In Section 2,
we prove that (\ref{oursolution}) satisfies the
three conditions ($a$) -- ($c$) and therefore gives
the Seiberg-Witten map. We also show that it is 
equivalent to (\ref{liu}) conjectured in \cite{Liu}. 
In Section 3, we discuss its relation to the
coupling of the noncommutative gauge field
to the Ramond-Ramond potentials in string theory.\footnote{In 
the course of this work, we were
informed of a work in progress by S. Das and N.V. Suryanarayana  
on some aspect of the Ramond-Ramond currents.}
In Section 4, we discuss applications and extensions of
our result.

\bigskip

After the first version of this paper appeared, we 
received two papers \cite{Mukhi2}, \cite{Liu2}, whose contents overlap with
Section 3 of this paper. 

\section{Proof}
\setcounter{equation}{0}

In this section, we will prove that (\ref{oursolution})
obeys the three conditions ($a$) -- ($c$) for the Seiberg-Witten
map. The gauge invariance ($a$) is manifest
because of the use of the open Wilson line 
\cite{Ishibashi:2000hs} -- \cite{Gross:2000ba}. 
We will show that it 
also satisfies the Bianchi identity (b) and
the initial condition (c). 

\subsection{Bianchi identity}

In order to prove the Bianchi identity, it is useful
to introduce the following  currents of rank $2s$,
\ber\label{currents}
&& J^{i_1\cdots i_{2s}}(k) = \frac{1}{{\rm Pf}(\theta)}
\int dx *\left[~~ e^{ikx}\int_0^1 d\tau_1 \left(\theta - 
\theta \hat{F}(x+l\tau_1) \theta\right)^{[i_1,i_2}
\cdots \right.
\cr
&& \qquad \left. \times
\int_0^1 d\tau_n \left(\theta - 
\theta \hat{F}(x+l\tau_n) \theta\right)^{i_{2s-1},i_{2s}]} 
 P\exp\left(i\int_0^1 \hat{A}_i(x+l\tau)
l^i d\tau \right)~\right].
\eer
Here the indices $i_1,\ldots,i_{2s}$ are totally antisymmetrized
with a factor of $1/(2s)!$ for each term.
For noncommutative gauge theory in $2n$ dimensions,
the Seiberg-Witten map (\ref{oursolution}) can be written
as
\eqn{rewritesw}
{ F_{ij}(k) + \theta_{ij}^{-1}\delta(k)
 = - {1 \over 2^{n-1} (n-1)!} 
\epsilon_{iji_1 \cdots i_{2n-2}}
J^{i_1 \cdots i_{2n-2}}(k).}
Therefore, to prove that the left-hand side of 
(\ref{rewritesw}) obeys the Bianchi identity, 
it is sufficient to show that these currents
are conserved,
\eqn{currentconserv}
{ k_{i_1} J^{i_1 \cdots i_{2s}}(k) = 0.}

The conservation law can be proven by performing
integration by parts in the $\tau$-integrals
in (\ref{currents}). Before describing a proof
for general $s$, it would be instructive to
show how it works for $s=1$ and $s=2$. 
When $s=1$, 
\ber \label{noncummutative-current-1}
&&k_i \int dx \ast \biggl[ e^{ikx}
\int_0^1 d \tau'
(\theta - \theta \hat{F} (x+ l \tau') \theta )^{ij}
P \exp \left(
i \int_0^1 \hat{A}_i ( x+l \tau) l^i d \tau \right)
\biggr] \cr
&&=
\int dx \ast \biggl[
e^{ikx} \int_0^1 d\tau'
(l^j - \theta^{ji} \hat{F}_{ii'} l^{i'})
P \exp \left(
i \int_0^1 \hat{A}_i ( x+l \tau) l^i d \tau \right)
\biggr] \cr
&& 
 =
i \theta^{jj'} \int dx~ \partial_{j'} \ast \biggl[
e^{ikx}
P \exp \left(
i \int_0^1 \hat{A}_i ( x+l \tau) l^i d \tau \right)
\biggr] \cr
&& = 0.
\eer
Here we decomposed the factor in the second line as follows:
\begin{equation}
l^j - \theta^{ji} \hat{F}_{ii'} l^{i'}
= i \theta^{jj'} (i k_{j'} + i \partial_{j'} \hat{A}_{i'} l^{i'})
+ \theta^{jj'} l^{i'} D_{i'} \hat{A}_{j'},
\end{equation}
and used the identity that
\begin{equation}
\int dx \ast \biggl[
e^{ikx} \int_0^1 d\tau'~
l^{i'} D_{i'} \hat{A}_{j'} (x + l \tau')
P \exp \left(
i \int_0^1 \hat{A}_i ( x+l \tau) l^i d \tau \right)
\biggr] =0,
\end{equation}
which was shown in (B.4) in \cite{Okawa:2001sh}.

To prove the current conservation for  
$s=2$, we use the following identity,
\begin{eqnarray}
&& \int dx \ast \biggl[ e^{ikx}
\int_0^1 d \tau_1~
l^i ( \hat{F}_{ij} (x+ l \tau_1)-  \theta^{-1}_{ij} )
\int_0^1 d \tau_2~
{\cal O}(x+ l \tau_2)
\nonumber \\ && \qquad \qquad \qquad \qquad \times
P \exp \left(
i \int_0^1 \hat{A}_i ( x+l \tau) l^i d \tau \right)
\biggr]
\nonumber \\ &=&
-i \int dx \ast \biggl[ e^{ikx}
\int_0^1 d \tau'
D_j {\cal O}(x+ l \tau')
P \exp \left(
i \int_0^1 \hat{A}_i ( x+l \tau) l^i d \tau \right)
\biggr].
\label{identity-for-conservation}
\end{eqnarray}
The conservation law for $s=2$,
\begin{eqnarray}
k_i \int dx \ast \biggl[ e^{ikx}
\int_0^1 d \tau_1
(\theta - \theta \hat{F} (x+ l \tau_1) \theta )^{i[j}
\int_0^1 d \tau_2
(\theta - \theta \hat{F} (x+ l \tau_2) \theta )^{k,l]}
\nonumber \\
\times P \exp \left(
i \int_0^1 \hat{A}_i ( x+l \tau) l^i d \tau \right)
\biggr] =0,
\label{noncommutative-current-2}
\end{eqnarray}
follows from this by setting
${\cal O}= \hat{F}_{kl} - \theta^{-1}_{kl}$
and using the Bianchi identity 
\eqn{noncommbianchi}
{D_j \hat{F}_{kl} + D_k \hat{F}_{lj} + D_l \hat{F}_{jk} =0}
for $\hat{F}$. 
What remains is to show (\ref{identity-for-conservation}).	
This follows from the follwoing two identities. The first one is
\begin{eqnarray}
&& \int dx \ast \biggl[ e^{ikx}
\int_0^1 d \tau_1~
l^i D_i \hat{A}_{j} (x+ l \tau_1) 
\int_0^1 d \tau_2~
{\cal O}(x+ l \tau_2)
P \exp \left(
i \int_0^1 \hat{A}_i ( x+l \tau) l^i d \tau \right)
\biggr]
\nonumber \\ &&=
\int dx \ast \biggl[ e^{ikx}
\int_0^1 d \tau'
[\hat{A}_j, {\cal O}] (x+ l \tau')
P \exp \left(
i \int_0^1 \hat{A}_i ( x+l \tau) l^i d \tau \right)
\biggr],
\label{identity-1}
\end{eqnarray}
which can be derived from (B.5) in \cite{Okawa:2001sh}.
The second one is
\begin{eqnarray}
&& \int dx \ast \biggl[ e^{ikx}
\int_0^1 d \tau_1~
\left\{ -k_j - l^i \partial_j \hat{A}_{i} (x+ l \tau_1) \right\}
\int_0^1 d \tau_2~
{\cal O}(x+ l \tau_2)
\nonumber \\ && \qquad \qquad \qquad \qquad \times
P \exp \left(
i \int_0^1 \hat{A}_i ( x+l \tau) l^i d \tau \right)
\biggr]
\nonumber \\ &=&
-i \int dx \ast \biggl[ e^{ikx}
\int_0^1 d \tau'
\partial_j {\cal O}(x+ l \tau')
P \exp \left(
i \int_0^1 \hat{A}_i ( x+l \tau) l^i d \tau \right)
\biggr],
\label{identity-2}
\end{eqnarray}
where we performed integration by parts on $\hat{A}_i$. 
By combining (\ref{identity-1}) and (\ref{identity-2})
using 
\begin{equation}
l^i D_i \hat{A}_j - l^i \partial_j \hat{A}_i - k_j
= l^i ( \hat{F}_{ij} - \theta^{-1}_{ij} ),
\label{identity-3}
\end{equation}
we obtain the identity (\ref{identity-for-conservation}).

To give a proof of the conservation law (\ref{currentconserv})
for general $s$, 
it is most convenient to use the Matrix Theory language
\cite{Banks:1997vh} -- \cite{Ishibashi:1997xs}. 
The noncommutative gauge theory with a commutative time coordinate $t$
and $2n$ noncommutative space coordinates $x^i$ ($i=1,\ldots,2n$) 
can be constructed from Matrix Theory by setting the matrix
variables $X^i$ in the form, 
\eqn{noncommbackground}
{X^i = x^i + \theta^{ij} \hat{A}_j(x),} 
where $x^i$ obeys the commutation relation,
\eqn{communagain}
{[x^i, x^j] = -i \theta^{ij}.}
Formulae in noncommutative gauge theory can then
be expressed in the Matrix Theory language according to
the map \cite{background},
\ber \label{ncymrule}
[X^i,X^j] &=& -i\left(\theta^{ij} - \theta^{ii'}\hat{F}_{i'j'}
\theta^{j'j}\right),\cr
e^{ikX} &=& *\left[ ~e^{ikx} 
P\exp\left(i\int_0^1 \hat{A}_i(x+l\tau) l^i \right)\right],\cr
{\tr}\left( \cdots \right) & = & 
\frac{1}{{\rm Pf}(\theta)} \int dx  *\left[ \cdots \right],
\eer
with $l^i = k_j \theta^{ji}$.
For a more precise description of the map between
gauge invariant operators of Matrix Theory
and the noncommutative gauge theory, see \cite{Okawa:2001if}.\footnote{
See also the formula (\ref{matrix-to-noncommutative})
given in Appendix.}

Following the rule (\ref{ncymrule}), we can 
 express the currents (\ref{currents}) in
noncommutative gauge theory 
using the Matrix Theory variables as
\ber\label{tauordered}
J(k) &=& {\rm tr}\left( e^{ikX} \right), \cr
J^{ij}(k) &=& i~ {\rm tr} \left(  [X^i, X^j]e^{ikX}\right), \cr
J^{ijlm}(k) &=& \frac{i^2}{3} \int_0^1 d\tau {\rm tr}
\left( [X^i,X^j] e^{i\tau kX}
[X^l, X^m] e^{i(1-\tau)kX} \right) \cr
&& + \frac{i^2}{3} \int_0^1 d\tau {\rm tr}
\left( [X^i,X^l] e^{i\tau kX}
[X^m, X^j] e^{i(1-\tau)kX} \right) \cr
&& + \frac{i^2}{3} \int_0^1 d\tau {\rm tr}
\left( [X^i,X^m] e^{i\tau kX}
[X^j, X^l] e^{i(1-\tau)kX} \right), \cr
&&\cr
&\vdots&\cr
&&\cr
J^{i_1\cdots i_{2n}}(k) &=& \frac{i^n (n-1)!}{(2n)!}
\int_0^1 d\tau_1 \int_{\tau_1}^1 d\tau_2
 \cdots \int_{\tau_{n-2}}^1d\tau_{n-1}
\cr
&& \times {\rm tr} \left(
 [X^{i_1}, X^{i_2}] e^{i\tau_1 kX}[X^{i_3},X^{i_4}]e^{i(\tau_2-\tau_1)kX}
\cdots [X^{i_{2n-1}}, X^{i_{2n}} ] 
e^{i(1-\tau_{n-1})kX}\right) \cr
&& \qquad +\left( ((2n)!-1)~{\rm more~terms~to~antisymmetrize~
the~indices}\right).\cr
&&\eer
This facilitates our proof of the conservation law,
\eqn{conserve}{
k_{i_1} J^{i_1 \cdots i_{2s}}(k) = 0.} 

In order to prove the conservation law in the Matrix Theory
language,  we will make use of the cyclicity 
of the trace, ${\rm tr}(AB)={\rm tr}(BA)$. A care is needed 
here since this does not necessarily hold for infinite
dimensional matrices. For example, in the background $X^i=x^i$ 
which gives rise to a noncommutative gauge theory from Matrix Theory, 
we have
\eqn{noncomm}
{[x^i, x^j] = -i\theta^{ij}.}
Therefore ${\rm tr}(x^ix^j)={\rm tr}(x^jx^i)$ is obviously untrue here. 
Fortunately, the conservation law  can be proven under 
the weaker assumption about the cyclicity of the trace as,
\eqn{assumption}
{{\rm tr}\left([X^i,X^j] {\cal O}\right)
={\rm tr}\left({\cal O}[X^i,X^j]\right),~~
{\rm tr}\left(e^{ikX}{\cal O}\right)
={\rm tr}\left({\cal O}e^{ikX}\right), }
for any ${\cal O}$ 
generated by any number of commutators
$[X^i,X^j]$ and exponentials $e^{ik'X}$
with a possibility of a single insertion of $X^i$. 
This holds for
$X^i$ considered in this paper.
($X^i = x^i + \theta^{ij} \hat{A}_j(x)$
and we are allowed to perform integration by parts
on $\hat{A}_j$.)
 
As a warm-up, let us repeat the proof for $s=1$ and $s=2$
using the Matrix Theory language.
For $s=1$, we can show the conservation
for matrices $X^i$ satisfying (\ref{assumption}) as follows:
\begin{eqnarray}
k_i J^{ij}(k) &=&
{\rm tr}\left( [ikX, X^j] e^{ikX} \right)
\nonumber \\
&=& \int_0^1 d \tau~ {\rm tr} \left(
e^{i \tau kX} [ikX,X^j] e^{i(1-\tau)kX} \right)
= {\rm tr} \left( [e^{ikX}, X^j] \right)
=0.
\label{checkone}
\end{eqnarray}
For $s=2$, we need to perform the integration by parts
in $\tau$ as
\begin{eqnarray}
&& k_i J^{ijlm}(k) \nonumber \\
&&= \frac{i}{3} \int_0^1 d\tau ~{\rm tr}
\left( e^{i(1-\tau)kX} [ikX,X^{j}] e^{i\tau kX}
[X^l, X^{m}]  \right)
+ (2~{\rm more~terms})\nonumber \\
&&= -\frac{i}{3} \int_0^1 d \tau \frac{d}{d\tau} ~{\rm tr}
\left( e^{i(1-\tau)kX} X^{j} e^{i\tau kX} [X^l, X^{m}] \right)
+ (2~{\rm more~terms})\nonumber \\
&&= -\frac{i}{3} {\rm tr}\left(X^{j} e^{ikX} [X^l, X^{m}]
- e^{ikX} X^{j} [X^l, X^{m}] \right)
+ (2~{\rm more~terms})\nonumber \\
&&= -\frac{i}{3} {\rm tr}\left([[X^l, X^{m}], X^{j}] e^{ikX} \right)
+ (2~{\rm more~terms})\nonumber \\
&&=0.
\label{checktwo}
\end{eqnarray}
To go from the fourth to 
the fifth line, we used the cyclicity of the trace. The
last line follows from the Jacobi identity. 

One can easily
see that each step in (\ref{checkone}) and (\ref{checktwo})
has a corresponding step in the proof
(\ref{noncummutative-current-1}) - (\ref{identity-3})
using the gauge theory variables. If one wishes,
one can also re-express the proof for arbitrary $s$
in the following using the gauge theory variables,
although the use of Matrix Theory variables
substantially simplifies the proof.

Now we are ready to prove the conservation law
for arbitrary $s$.\footnote{
The following proof also resolves 
the question raised 
in \cite{Schwarz:2001ps} regarding the gauge invariance of the 
Ramond-Ramond couplings and extends the earlier
work \cite{VanRaamsdonk:1999in} on conservation of currents
in Matrix Theory.}
In the original form of the current in (\ref{tauordered}),
the indices $i_1, i_2 ,\ldots, i_{2s}$
are totally antisymmetrized. However, we can always bring
one of them $i_1$ to the first using the cyclic symmetry
of the $\tau$-integral form,
while the rest of the indices $i_2, i_3, \ldots i_{2n}$ are
still totally antisymmetrized.
One of the terms appeared in (\ref{conserve})
is then
\begin{eqnarray}
&& -i \int_0^1 d \tau_1 \int_{\tau_1}^1 d \tau_2 \cdots
\int_{\tau_{n-2}}^1 d \tau_{n-1} ~{\rm tr}~
e^{i (1-\tau_{n-1}) kX} [ ikX, X^{i_2}] e^{i \tau_1 kX}
\nonumber \\ && \qquad \times
[X^{i_3}, X^{i_4}] e^{i (\tau_2 - \tau_1) kX} 
[X^{i_5}, X^{i_6}] e^{i (\tau_3 - \tau_2) kX} \cdots
[X^{i_{2n-1}}, X^{i_{2n}}]
\nonumber \\ &=&
i \int_0^1 d \tau_1 \int_{\tau_1}^1 d \tau_2 \cdots
\int_{\tau_{n-2}}^1 d \tau_{n-1}
\left( \frac{d}{d \tau_1} + \frac{d}{d \tau_2} + \cdots
+ \frac{d}{d \tau_{n-1}} \right)\cr
&&~~~\times
{\rm tr}~ e^{i (1-\tau_{n-1}) kX} X^{i_2} e^{i \tau_1 kX}
\nonumber \\ && \qquad \times
[X^{i_3}, X^{i_4}] e^{i (\tau_2 - \tau_1) kX}
[X^{i_5}, X^{i_6}] e^{i (\tau_3 - \tau_2) kX} \cdots
[X^{i_{2n-1}}, X^{i_{2n}}]
\nonumber \\ &=&
-i \int_0^1 d \tau_2 \int_{\tau_2}^1 d \tau_3 \cdots
\int_{\tau_{n-2}}^1 d \tau_{n-1}
~{\rm tr}~ e^{i (1-\tau_{n-1}) kX} X^{i_2}
\nonumber \\ && \qquad \times
[X^{i_3}, X^{i_4}] e^{i \tau_2 kX}
[X^{i_5}, X^{i_6}] e^{i (\tau_3 - \tau_2) kX} \cdots
[X^{i_{2n-1}}, X^{i_{2n}}]
\nonumber \\ &&
+i \int_0^1 d \tau_1 \int_{\tau_1}^1 d \tau_2 \cdots
\int_{\tau_{n-3}}^1 d \tau_{n-2}
~{\rm tr}~ X^{i_2} e^{i \tau_1 kX}
\nonumber \\ && \qquad \times
[X^{i_3}, X^{i_4}] e^{i (\tau_2 - \tau_1) kX} \cdots
e^{(1 -\tau_{n-2})} [X^{i_{2n-1}}, X^{i_{2n}}].
\label{before-Jacobi}
\end{eqnarray}
In the last step, we used the formula
derived using the integration by parts,
\begin{eqnarray}
&& \int_0^1 d \tau_1 \int_{\tau_1}^1 d \tau_2 \cdots
\int_{\tau_{n-2}}^1 d \tau_{n-1}
\left( \frac{d}{d \tau_1} + \frac{d}{d \tau_2} + \cdots
+ \frac{d}{d \tau_{n-1}} \right)
f(\tau_1, \tau_2, \ldots, \tau_{n-1})
\nonumber \\ &=&
- \int_0^1 d \tau_2 \int_{\tau_2}^1 d \tau_3 \cdots
\int_{\tau_{n-2}}^1 d \tau_{n-1}~
f(0, \tau_2, \ldots, \tau_{n-1})
\nonumber \\ &&
+ \int_0^1 d \tau_1 \int_{\tau_1}^1 d \tau_2 \cdots
\int_{\tau_{n-3}}^1 d \tau_{n-2}~
f(\tau_1, \tau_2, \ldots, \tau_{n-2}, 1),
\end{eqnarray}
where $f$ is an arbitrary function of 
$\tau_1, \tau_2, \ldots, \tau_{n-1}$.
Using the antisymmetry in the indices,
$i_2, i_3, \ldots, i_{2n}$,
we can rewrite (\ref{before-Jacobi}) as follows:
\begin{eqnarray}
&& -i \int_0^1 d \tau_2 \int_{\tau_2}^1 d \tau_3 \cdots
\int_{\tau_{n-2}}^1 d \tau_{n-1}
~{\rm tr}~ [X^{i_2}, [X^{i_3}, X^{i_4}] ]
\nonumber \\ && \qquad \times
e^{i \tau_2 kX}
[X^{i_5}, X^{i_6}] e^{i (\tau_3 - \tau_2) kX} \cdots
[X^{i_{2n-1}}, X^{i_{2n}}] e^{i (1-\tau_{n-1}) kX}.
\end{eqnarray}
This vanishes because of the Jacobi identity.

We have proven the conservation of the current
(\ref{currents}). Thus (\ref{rewritesw})
satisfies the Bianchi identity.

\subsection{Initial condition}

Since the conditions ($a$) and ($b$) are equivalent
to the Seiberg-Witten equation (\ref{sw}), we now have  
a solution to the equation, modulo field redefinition
and gauge transformation. What remains to verify is
the initial condition ($c$). Although we can check this
directly by expanding the map (\ref{oursolution})
in powers of $\theta$, it is more useful to rewrite (\ref{oursolution})
in such a way that the initial condition is manifest. 
In this process, we find that (\ref{oursolution}) is equivalent to
(\ref{liu}), therefore proving the conjecture in \cite{Liu}. 

To see the relation between (\ref{oursolution}) and (\ref{liu}), 
let us first show the identity
\ber \label{firstidentity}
&& \frac{1}{{\rm Pf}(\theta)}
(\theta - \theta\hat{f}\theta)^{n-1}_{ij}
-\sqrt{\det(1-\hat{f}\theta)} \left({1 \over 1 -\hat{f}\theta}
 \hat{f}\right)_{ij}\cr
&&
= \frac{1}{{\rm Pf}(\theta)} \theta^{-1}_{ij}
{\rm Pf}(\theta - \theta \hat{f} \theta).
\eer
This can be shown by writing the two terms on the left-hand
side of the equation as
\eqn{first}{\frac{1}{{\rm Pf}(\theta)}
(\theta - \theta\hat{f}\theta)^{n-1}_{ij}
= \frac{1}{{\rm Pf}(\theta)} {\rm Pf}(\theta - \theta \hat{f} \theta)
  \left({1 \over \theta - \theta \hat{f} \theta}\right)_{ij},}
and\footnote{We define the sign of the square root,
$\sqrt{\det(1-\hat{f}\theta)}$, so that it agrees with
that of ${\rm Pf}(\theta - \theta \hat{f}\theta)/{\rm Pf}(\theta)$.}
\eqn{second}
{\sqrt{\det(1-\hat{f}\theta)} \left({1 \over 1 -\hat{f}\theta}
 \hat{f}\right)_{ij}
 = \frac{1}{{\rm Pf}(\theta)}
{\rm Pf}(\theta - \theta \hat{f} \theta)
    \left({1 \over \theta - \theta \hat{f} \theta}\theta \hat{f}
\right)_{ij},}
and taking the difference of the two.
Therefore we find
\ber \label{difference}
&& \frac{1}{{\rm Pf}(\theta)} \int dx *\left[ e^{ikx}
(\theta - \theta\hat{f}\theta)^{n-1}_{ij}
P\exp\left(i\int_0^1 \hat{A}_i(x+l\tau)l^i d\tau\right)
\right]\cr
&=&\int dx *\left[e^{ikx}
 \sqrt{\det(1-\hat{f}\theta)} \left({1 \over 1 -\hat{f}\theta}
 \hat{f}\right)_{ij}P\exp\left(i\int_0^1 \hat{A}_i(x+l\tau)l^i d\tau\right)
\right]\cr 
&&~~+ \theta^{-1}_{ij} 
\frac{1}{{\rm Pf}(\theta)} \int dx *\left[
e^{ikx} {\rm Pf}(\theta - \theta \hat{f} \theta)
P\exp\left(i\int_0^1 
   \hat{A}_i(x+l\tau) l^i d\tau\right) \right].
\eer

Next we show
\begin{equation}
\frac{1}{{\rm Pf}(\theta)} \int dx *\left[
e^{ikx} {\rm Pf}(\theta - \theta \hat{f} \theta)
P\exp\left(i\int_0^1 
   \hat{A}_i(x+l\tau) l^i d\tau\right) \right]
= \delta(k).
\label{maximumrr}
\end{equation}
Note that the left-hand side is the Ramond-Ramond
current of the maximum rank $2n$,
\ber \label{same}
&& \frac{1}{{\rm Pf}(\theta)} \int dx *\left[
e^{ikx} {\rm Pf}(\theta - \theta \hat{f} \theta)
P\exp\left(i\int_0^1 
   \hat{A}_i(x+l\tau) l^i d\tau\right) \right]\cr
&=& \frac{1}{2^n n!}
\epsilon_{i_1\cdots i_{2n}} J^{i_1\cdots i_{2n}}(k).
\eer
To prove (\ref{maximumrr}), 
it is simplest to use the Matrix Theory representation 
(\ref{tauordered}). We will show the
current $J^{i_1\cdots i_{2n}}(k)$ of the maximum rank\footnote{In 
this paper, we are setting all the scalar fields
to be zero. Thus  $J^{i_1\cdots i_{2n}}(k)$
is the current of the maximum rank
for the noncommutative gauge theory in $(2n+1)$
dimensions.}
is invariant under an arbitrary infinitesimal variation of the matrix 
variable near the background $X^i = x^i$ with $[x^i,x^j] = -i \theta^{ij}$,
namely, it is topological. Once it is shown, 
we can evaluate the left-hand side of (\ref{maximumrr})
at the background $X^i = x^i$ which corresponds to $\hat{A}_i (x)=0$
and find
\begin{eqnarray}
&& \frac{1}{{\rm Pf}(\theta)} \int dx *\left[
e^{ikx} {\rm Pf}(\theta - \theta \hat{f} \theta)
P\exp\left(i\int_0^1 
   \hat{A}_i(x+l\tau) l^i d\tau\right) \right]
\nonumber \\
&=& \frac{1}{{\rm Pf}(\theta)} \int dx *\left[
e^{ikx} {\rm Pf}(\theta) \right]
= \delta(k).
\label{zerobackground}
\end{eqnarray}

Now let us prove that the right-hand side of (\ref{same}) is
indeed topological. 
It is instructive to consider the simplest case of $n=1$ first,  
\ber \label{firstone}
&&\delta{\rm tr}\left(\epsilon_{ij} [X^i, X^j] e^{ikX}\right)\cr
&&= \epsilon_{ij} ~{\rm tr}
\left(~ 2  [\delta X^i , X^j ] e^{ikX}
                  +\int_0^1 d\tau [X^i,X^j] e^{i\tau kX}
                        ik_m \delta X^m e^{i(1-\tau) kX}\right) \cr
&& =  (2 \epsilon_{ij}
ik_m   
+ ik_i\epsilon_{jm}){\rm tr}\left(
    \delta X^i \int_0^1 d\tau e^{i\tau kX}
[X^j,X^m] e^{i(1-\tau) kX}\right) \cr
&& = 0.
\eer
To go from the second to the third line, we used the cyclicity
of the trace. In the last line, we
used the identity in two dimensions,
\eqn{twodimid}
{2 \epsilon_{ij} \epsilon^{jl} + \epsilon_{jm} \epsilon^{jm}
\delta_{i}^{l} = 0.}
In general, we have
\ber \label{secondone}
&&\epsilon_{i_1\cdots i_{2n}}
\int_0^1 d \tau_1 \int_{\tau_1}^1 d \tau_2 \cdots
\int_{\tau_{n-2}}^1 d \tau_{n-1}~
\delta{\rm tr}\left( 
 [X^{i_1}, X^{i_2}] e^{i\tau_1 kX}\cdots [X^{i_{2n-1}}, X^{i_{2n}}]
e^{i(1-\tau_{n-1})kX} \right) \cr
&& =  ( 2n  k_{i_{2n}}\epsilon_{i i_1\cdots i_{2n-1}}
+ k_i \epsilon_{i_1 i_2\cdots i_{2n}}) 
\epsilon^{i_1\cdots i_{2n}}
\int_0^1 d \tau_0 \int_{\tau_0}^1 d \tau_1 \cdots
\int_{\tau_{n-2}}^1 d \tau_{n-1}\cr
&&~~~~~\times {n
\over (2n)!}\epsilon_{i_1'\cdots i_{2n}'}{\rm tr} \left(
\delta X^i e^{i\tau_0kX}[X^{i_1'}, X^{i_2'}] 
e^{i(\tau_0-\tau_1)kX}\cdots [X^{i_{2n-1}'}, X^{i_{2n}'}]
e^{i(1-\tau_{n-1})kX} \right) \cr
&&=0.
\eer
In the last line, we used the identity in $2n$ dimensions,
\eqn{twondimensions}
{ 2n \epsilon_{i i_1\cdots i_{2n-1}} \epsilon^{i_1 \cdots i_{2n-1}j}
 + \epsilon_{i_1 \cdots i_{2n}}\epsilon^{i_1 \cdots i_{2n}} 
\delta_i^j = 0.}
Thus we have proven
that the right-hand side of (\ref{same}) is topological.
Combining (\ref{difference}) and (\ref{zerobackground}),
we find 
\ber \label{finally}
&&
\frac{1}{{\rm Pf}(\theta)} \int dx *\left[ e^{ikx}
(\theta - \theta\hat{f}\theta)^{n-1}_{ij}
P\exp\left(i\int_0^1 
   \hat{A}_i(x+l\tau) l^i d\tau\right) \right]\cr
&&-
\int dx *\left[ e^{ikx}
\sqrt{\det(1-\hat{f}\theta)} \left({1 \over 1 -\hat{f}\theta}
 \hat{f}\right)_{ij}P\exp\left(i\int_0^1 
   \hat{A}_i(x+l\tau) l^i d\tau\right) \right] \cr
&&=\theta^{-1}_{ij}\delta(k).
\eer
Therefore, the conjectured expression (\ref{liu})
agrees with (\ref{oursolution}).
This completes the proof that (\ref{oursolution})
gives an exact Seiberg-Witten map.

\section{Relation to the Ramond-Ramond coupling}
\setcounter{equation}{0}

In Section 2, we have proven that (\ref{oursolution}) satisfies
the conditions ($a$) -- ($c$).
Now we would like to explain the string theoretical
origin of the formula.
As we mentioned in Introduction, we found the
expression for the Seiberg-Witten map (\ref{oursolution}) 
by studying the coupling of noncommutative gauge
theory realized on $p$-branes to the Ramond-Ramond $(p-1)$-form
in the bulk. The dual of the Ramond-Ramond current 
$J^{i_1\cdots i_{p-1}}$ on the $(p+1)$-dimensional 
worldvolume is a two-form.
It is clear that this two-form must be invariant
under the noncommutative gauge transformation,
and thus it obeys the condition ($a$). The condition ($b$) is satisfied
since the coupling should also be invariant under the
Ramond-Ramond gauge transformation 
$C^{(p-1)} \rightarrow C^{(p-1)} + d\epsilon$.
If we assume that there is no $\alpha'$ corrections to the
Ramond-Ramond coupling, we can expect that the two-form is
related to the field strength $F_{ij}$ of the commutative
variable as follows \cite{Polchinski:1988tu} -- \cite{Douglas:1995bn}:
\begin{equation}
\int C^{(p-1)} \wedge ( F + \theta^{-1} ).
\end{equation}
If that is the case, the condition ($c$) should also
hold. This
was our motivation for (\ref{oursolution}).

The couplings of noncommutative 
gauge theory to closed string states in the bulk
can be derived in various different ways. 
One approach is to evaluate disk amplitudes on a D$p$ brane with
a background of NS-NS two-form field and take the Seiberg-Witten limit. 
In \cite{Okawa:2001sh},
the energy-momentum tensor of the noncommutative theory was 
derived in this way. 
Alternatively, one can start with Matrix Theory \cite{Banks:1997vh}
($i.e.$,  many D0 branes instead of a D$p$ brane), 
compute the coupling of
the bulk fields to the matrix variables, and evaluate it in the background 
 which gives rise to the noncommutative gauge theory on a D$p$ 
brane \cite{Banks:1997vh} -- \cite{Ishibashi:1997xs}.
This approach was suggested in \cite{Das:2001ur}
and was carried out explicitly 
in \cite{Okawa:2001if} in the case of the coupling to the bulk graviton,  
where it was found to give the same result as that obtained using
the first approach \cite{Okawa:2001sh}. 

Here we will adopt the second approach since the currents couple
to the Ramond-Ramond potentials have already been studied in 
Matrix Theory \cite{Taylor:1999gq} -- \cite{Myers:1999ps}.\footnote
{In the case of constant Ramond-Ramond potentials,
the coupling to noncommutative gauge fields was studied in
\cite{Mukhi:2000zm}.}
For our purpose, it is sufficient to have the currents coupled to the 
space-time components of the Ramond-Ramond potential,
$C_{0i_1\cdots i_p}$, where the index $0$
is for the timelike coordinate 
and $i=1,\ldots,9$ are for the spacelike coordinates in the IIA string 
theory. The relevant couplings deduced in \cite{Taylor:1999gq} -- 
\cite{Myers:1999ps} are of the form,
\eqn{matrixrr}{
\int dt~ {\rm Str}\left( C_0(t,X)
+ C_{0ij}(t,X) [X^i,X^j] + C_{0ijkl}(t,X) [X^i,X^j][X^k,X^l] 
+ \cdots \right).}
Here $X^i$ are matrix coordinates, and the symmetrized trace 
${\rm Str}$ is defined by expanding $C_{0i_1\cdots i_p}(t,X)$ 
in powers of $X$'s
and totally symmetrize them together with $[X^i,X^j]$, each of which
is treated as one unit in the symmetrization. 
In the momentum basis, the currents coupled to  
the Ramond-Ramond potentials can be read off from (\ref{matrixrr}) as 
\ber \label{rrsource}
J(k) &=& {\rm Str}\left( e^{ikX} \right), \cr
J^{ij}(k) &=&
i~{\rm Str} \left(  [X^i, X^j]e^{ikX}\right), \cr
&&\cr
          & \vdots & \cr
&&\cr
J^{i_1\cdots i_{2s}}(k) & = & \frac{i^s}{(2s)!}{\rm Str} \left(
 [X^{i_1}, X^{i_2}] \cdots [X^{i_{2s-1}}, X^{i_{2s}} ] 
e^{ikX}\right) \cr
&&+\left( ((2s)!-1)~{\rm more~terms~to~antisymmetrize~
the~indices}\right).
\eer

We should point out that the symmetrized traces in (\ref{matrixrr})
and (\ref{rrsource}) make sense only when $X$'s are
trace class operators since they are defined by expanding
$C_{0i_1\cdots i_p}(X)$ in powers of $X$'s before taking
the trace. If $X$'s are infinite dimensional, a trace of 
powers of $X$'s may not be well-defined, though a trace of
$e^{ikX}$ may still exist. In fact, this is the case when  
$X^i = x^i + \theta^{ij} \hat{A}_j(x)$ with $[x^i, x^j ] 
= -i\theta^{ij}$. 

On the other hand, the currents (\ref{tauordered}) we used 
in Section 2 make sense even when $X$'s are of the form 
$X^i = x^i + \theta^{ij} \hat{A}_j(x)$. Moreover, they
agree with (\ref{rrsource}) when $X$'s are finite 
dimensional. In fact, if ${\cal O}_1 \cdots {\cal O}_n$
and $X$ are trace class operators, one can prove
\ber \label{equivalence}
 && {\rm Str}\left( {\cal O}_1 \cdots {\cal O}_n
e^{ikX} \right)  \cr
&& = \int_0^1 d\tau_1 \int_{\tau_1}^1 d\tau_2
\cdots \int_{\tau_{n-2}}^1 d\tau_{n-1}
{\rm tr} \left( {\cal O}_1 e^{i\tau_1 kX} {\cal O}_2
\cdots {\cal O}_n e^{i(1-\tau_{n-1})kX}\right) \cr
&&~~~~~~+\left(((n-1)!-1)~{\rm more~terms~to~symmetrize~}{\cal O}_i{\rm 's}
\right).
\eer
Here the symmetrized trace ${\rm Str}$ on the left-hand side
is defined by expanding $e^{ikX}$ in powers of $X$'s and symmetrizing
them with ${\cal O}_1 \cdots {\cal O}_n$. On
the other hand, the symmetrization on the right-hand side
exchanges  ${\cal O}_1 \cdots {\cal O}_n$ only. 
The equivalence (\ref{equivalence}) for $n=2$ has been proven 
in our previous paper \cite{Okawa:2001if}. A general proof
for arbitrary $n$ is given in Appendix of this paper. 
As shown in \cite{Okawa:2001if}, the $\tau$-integral
expressions such as (\ref{tauordered}) naturally arise from
disk amplitudes of a single closed string state and
an arbitrary number of open string states. In this
case, $\tau_1, \ldots, \tau_n$ are identified as 
locations of the open string vertex operators on
the boundary of the worldsheet disk. In \cite{Okawa:2001if},
this is shown explicitly for the coupling of $X$'s
to the graviton in the bulk. We expect the situation
is the same for the coupling to the Ramond-Ramond 
potentials. This is the string theory origin of
the formula (\ref{currents}).

\section{Discussion}

In this paper, we proved that (\ref{oursolution}) 
satisfies the conditions ($a$)
-- ($c$) for the Seiberg-Witten map. We
also showed that 
it is equivalent to (\ref{liu}) and
therefore proved the conjecture in \cite{Liu}.
 
The exact Seiberg-Witten map can be used to understand
the relation between the commutative and noncommutative
descriptions of D-branes with a strong NS-NS two-form field.
For example, it may be possible to study the noncommutative
solitons \cite{harvey} in the language of the 
commutative variables. 

In this paper, we set all the scalar fields to be zero
and focused on the Seiberg-Witten map between 
$A_i(x)$ and $\hat{A}_i(x)$. It is straightforward 
to include these in the analysis. We can also add
commutative dimensions by starting from many D$p$ branes
with $p > 0$ rather than D$0$ branes and by using
the results in \cite{Taylor:2000pr} and \cite{Myers:1999ps}
about the Ramond-Ramond coupling of these branes. 

\section*{Acknowledgements}
We would like to thank John Schwarz,
Mark Wise and Edward Witten for discussions.
We would also like to thank the referee
of this paper for useful comments. 
H.O. thanks the Institute for Theoretical
Physics, Santa Barbara for hospitality. 

The research was supported in part by
the DOE grant DE-FG03-92ER40701
and the Caltech Discovery Fund.
H.O. was also supported in part by
the NSF grant PHY99-07949.

\newpage

%%%%%%%%%% Appendices %%%%%%%%%%
\appendix
\renewcommand{\thesection}{Appendix \Alph{section}.}
\renewcommand{\theequation}{\Alph{section}.\arabic{equation}}

%%%%% Appendix A %%%%%
\section{$\tau$-ordered integral $=$ symmetrized trace}
\setcounter{equation}{0}

In this appendix, we will prove the equivalence
(\ref{equivalence}) of the symmetrized trace
and the $\tau$-ordered trace.\footnote{We assume
that the symmetrized trace is well-defined.
This means that, if we define the symmetrized trace
in terms of a power series expansion in $X$'s,
$X$ must be trace class operators.} 

First let us perform the $\tau$ integrals in 
(\ref{equivalence}) explicitly.
For any operators ${\cal O}_1, {\cal O}_2 , \ldots , {\cal O}_m$,
\begin{eqnarray}
&& \int_0^1 d \tau_1 \int_{\tau_1}^1 d \tau_2 \cdots
\int_{\tau_{m-2}}^1 d \tau_{m-1}
~{\rm tr}~ {\cal O}_1 e^{i \tau_1 kX}
{\cal O}_2 e^{i(\tau_2 -\tau_1) kX} \cdots
\nonumber \\ && \qquad \qquad \qquad \qquad \qquad \qquad \times
{\cal O}_{m-1} e^{i(\tau_{m-1} -\tau_{m-2}) kX}
{\cal O}_m e^{i(1 -\tau_{m-1}) kX}
\nonumber \\ &=&
\int_0^1 d \tau'_1 \int_0^{1-\tau'_1} d \tau'_2 
\int_0^{1-\tau'_1 -\tau'_2} d \tau'_3 \cdots
\int_0^{1-\tau'_1- \cdots -\tau'_{m-2}} d \tau'_{m-1}
~{\rm tr}~ {\cal O}_1 e^{i \tau'_1 kX}
{\cal O}_2 e^{i \tau'_2 kX} \cdots
\nonumber \\ && \qquad \qquad \qquad \qquad \qquad \qquad \qquad \times
{\cal O}_{m-1} e^{i \tau'_{m-1} kX}
{\cal O}_m e^{i(1 -\tau'_1 -\tau'_2 - \cdots -\tau'_{m-1}) kX}
\nonumber \\ &=&
\sum_{a_1 =0}^\infty \sum_{a_2 =0}^\infty \cdots
\sum_{a_m =0}^\infty 
\int_0^1 d \tau'_1 \int_0^{1-\tau'_1} d \tau'_2 \cdots
\int_0^{1-\tau'_1- \cdots -\tau'_{m-2}} d \tau'_{m-1}
\frac{1}{a_1 ! a_2 ! \cdots a_m !} \nonumber \\
&&\times
~{\rm tr}~ {\cal O}_1 (i \tau'_1 kX)^{a_1}
{\cal O}_2 (i \tau'_2 kX)^{a_2} \cdots
{\cal O}_{m-1} (i \tau'_{m-1} kX)^{a_{m-1}}\nonumber \\
&&~~~~~~~~\times
{\cal O}_m (i(1 -\tau'_1 -\tau'_2 - \cdots -\tau'_{m-1}) kX)^{a_m}
\nonumber \\ &=&
\sum_{a_1 =0}^\infty \sum_{a_2 =0}^\infty \cdots
\sum_{a_m =0}^\infty 
\frac{1}{(a_1 + a_2 + \cdots + a_m + m-1)!} \nonumber \\
&&~\times{\rm tr}~ {\cal O}_1 (ikX)^{a_1}
{\cal O}_2 (ikX)^{a_2} \cdots
{\cal O}_{m-1} (ikX)^{a_{m-1}}
{\cal O}_m (ikX)^{a_m}
\nonumber \\ &=&
\sum_{n=0}^\infty \frac{1}{(n+m-1)!}
\sum_{p_1 =0}^n \sum_{p_2 =0}^{n-p_1}
\sum_{p_3 =0}^{n-p_1-p_2} \cdots
\sum_{p_{m-1} =0}^{n-p_1-p_2- \cdots -p_{m-2}} \nonumber \\
&&~~\times{\rm tr}~ {\cal O}_1 (ikX)^{p_1}
{\cal O}_2 (ikX)^{p_2} \cdots
{\cal O}_{m-1} (ikX)^{p_{m-1}}
{\cal O}_m (ikX)^{n-p_1-p_2- \cdots p_{m-1}}.
\label{tau-integration}
\end{eqnarray}
Here we changed the integration variables in the first step as
\begin{equation}
\tau'_1 = \tau_1, \quad \tau'_2 = \tau_2 - \tau_1, \quad \cdots
\quad \tau'_{m-1} = \tau_{m-1} - \tau_{m-2},
\end{equation}
and used the following formula in performing the $\tau'$ integrals:
\begin{eqnarray}
&& \int_0^1 d \tau'_1 \int_0^{1-\tau'_1} d \tau'_2 \cdots
\int_0^{1-\tau'_1- \cdots -\tau'_{m-2}} d \tau'_{m-1}
\nonumber \\ && \times
{\tau'_1}^{\alpha_1 -1} {\tau'_2}^{\alpha_2 -2} \cdots
{\tau'_{m-1}}^{\alpha_{m-1} -1}
(1 -\tau'_1 -\tau'_2 - \cdots -\tau'_{m-1})^{\beta -1}
\nonumber \\
&=& \frac{\Gamma(\alpha_1) \Gamma(\alpha_2)
\cdots \Gamma(\alpha_{m-1}) \Gamma(\beta)}
{\Gamma(\alpha_1 + \alpha_2 + \cdots + \alpha_{m-1} + \beta)}
\quad {\rm for} \quad
\alpha_1, \alpha_2, \ldots, \alpha_{m-1}, \beta >0.
\end{eqnarray}
On the other hand, since
\begin{eqnarray}
&& {\rm Str} \left[ (ikX)^n {\cal O}_1 {\cal O}_2
\cdots {\cal O}_m \right]
\nonumber \\
&=& \frac{1}{(n+m-1)!}
\sum_{p_1 =0}^n \sum_{p_2 =0}^{n-p_1}
\sum_{p_3 =0}^{n-p_1-p_2} \cdots
\sum_{p_{m-1} =0}^{n-p_1-p_2- \cdots -p_{m-2}} 
n!~ {\rm tr} {\cal O}_1 (ikX)^{p_1}
{\cal O}_2 (ikX)^{p_2} \cdots
\nonumber \\ && \qquad \qquad \qquad \qquad \times
{\cal O}_{m-1} (ikX)^{p_{m-1}}
{\cal O}_m (ikX)^{n-p_1-p_2- \cdots p_{m-1}}
\nonumber \\ &&
+ \left( ((m-1)!-1) {\rm ~more~terms~to~symmetrize~in~}
{\cal O}_2, {\cal O}_3, \ldots, {\cal O}_m\right),
\end{eqnarray}
where {\rm Str} is the symmetrized trace with respect to
$X$, ${\cal O}_1$, \ldots, ${\cal O}_m$, we can write
\begin{eqnarray}
&& {\rm Str} \left[ e^{ikX} {\cal O}_1 {\cal O}_2
\cdots {\cal O}_m \right]
= \sum_{n=0}^\infty \frac{1}{n!}
{\rm Str} \left[ (ikX)^n {\cal O}_1 {\cal O}_2
\cdots {\cal O}_m \right]
\nonumber \\
&&= \int_0^1 d \tau_1 \int_{\tau_1}^1 d \tau_2 \cdots
\int_{\tau_{m-2}}^1 d \tau_{m-1}
{\rm tr} {\cal O}_1 e^{i \tau_1 kX}
{\cal O}_2 e^{i(\tau_2 -\tau_1) kX} \cdots
\nonumber \\ && \qquad \qquad \qquad \qquad \times
{\cal O}_{m-1} e^{i(\tau_{m-1} -\tau_{m-2}) kX}
{\cal O}_m e^{i(1 -\tau_{m-1}) kX}
\nonumber \\ \qquad \qquad
&& ~~~~+ \left(((m-1)!-1) {\rm ~more~terms~to~symmetrize}
\right)
\end{eqnarray}
Here we made use of the cyclicity of the trace.
Therefore, we have shown the equivalence of the two expressions
(\ref{rrsource}) and (\ref{tauordered}).
Furthermore, we can show that
\begin{eqnarray}
&& {\rm tr} \left[ \exp \left( i \int_0^1 d \tau k X \right)
\int_0^1 d \tau_1 {\cal O}_1 \cdots \int_0^1 d \tau_m {\cal O}_m
\right]
\nonumber \\
&&= \int_0^1 d \tau_1 \int_{\tau_1}^1 d \tau_2 \cdots
\int_{\tau_{m-2}}^1 d \tau_{m-1}
{\rm tr} {\cal O}_1 e^{i \tau_1 kX}
{\cal O}_2 e^{i(\tau_2 -\tau_1) kX} \cdots
\nonumber \\ && \qquad \qquad \qquad \qquad \times
{\cal O}_{m-1} e^{i(\tau_{m-1} -\tau_{m-2}) kX}
{\cal O}_m e^{i(1 -\tau_{m-1}) kX}
\nonumber \\ \qquad \qquad
&& ~~~~~+\left( ((m-1)!-1) {\rm ~more~terms~to~symmetrize}
\right)
\nonumber \\
&&= {\rm Str} \left[ e^{ikX} {\cal O}_1 {\cal O}_2
\cdots {\cal O}_m \right],
\label{matrix-to-noncommutative}
\end{eqnarray}
where the operators are $\tau$-ordered in the first line.
This formula is the generalization of (27)
in \cite{Okawa:2001if} to the case where more than two operators
are inserted and useful when we transform
the current $J^{i_1 i_2 \cdots i_{2n}}$ (\ref{tauordered})
to the form (\ref{currents}) used in the noncommutative
gauge theory.
\newpage
%%%%%%%%%% References %%%%%%%%%%

\renewcommand{\baselinestretch}{0.87}

%\bibliography{draft}
%\bibliographystyle{ssg}
\begingroup\raggedright\endgroup
\end{document}